\documentclass[12pt]{article}

\usepackage{amsmath}
\usepackage{amssymb}
\usepackage{amsfonts}
\usepackage{latexsym}
\usepackage{color}

\catcode `\@=11 \@addtoreset{equation}{section}

\catcode `\@=12

\newcommand{\be}{\begin{equation}}
\newcommand{\en}{\end{equation}}
\newcommand{\bea}{\begin{eqnarray}}
\newcommand{\ena}{\end{eqnarray}}
\newcommand{\beano}{\begin{eqnarray*}}
\newcommand{\enano}{\end{eqnarray*}}
\newcommand{\bee}{\begin{enumerate}}
\newcommand{\ene}{\end{enumerate}}

\newcommand{\N}{\mathfrak N}

\newcommand{\mc}{\mathcal}

\newcommand{\D}{{\mc D}}

\newcommand{\E}{{\cal E}}
\newcommand{\F}{{\cal F}}
\newcommand{\G}{{\cal G}}

\newcommand{\Lc}{{\cal L}}

\newcommand{\1}{1 \!\! 1}

\newcommand{\Hil}{\mc H}

\newtheorem{thm}{Theorem}
\newtheorem{cor}[thm]{Corollary}
\newtheorem{lemma}[thm]{Lemma}

\newenvironment{proof}{\noindent {\bf Proof --}}{\hfill$\square$ \vspace{3mm}\endtrivlist}

\catcode `\@=11 \@addtoreset{equation}{section}
\catcode `\@=12

\begin{document}

\thispagestyle{empty}

\vspace*{2cm}

\begin{center}
{\Large \bf Pseudo-bosons, Riesz bases and coherent states}   \vspace{2cm}\\

{\large F. Bagarello}\\
  Dipartimento di Metodi e Modelli Matematici,
Facolt\`a di Ingegneria,\\ Universit\`a di Palermo, I-90128  Palermo, Italy\\
e-mail: bagarell@unipa.it

\end{center}

\vspace*{2cm}

\begin{abstract}
\noindent In a recent paper, Trifonov suggested a possible explicit model of a PT-symmetric system based on a modification of the canonical commutation relation. Although being rather intriguing, in his treatment many mathematical aspects of the model have just been neglected, making most of the results of that paper purely formal. For this reason we are re-considering the same model and we repeat and extend  the same construction paying particular attention to all the subtle mathematical points. From our analysis the crucial role of Riesz bases clearly emerges. We also consider coherent states associated to the model.

\end{abstract}

\vspace{2cm}


\vfill


\newpage

\section{Introduction}

In a recent paper, \cite{tri}, Trifonov suggested a possible explicit model of a PT-symmetric system based on a modification of the canonical commutation relation (CCR). The physical relevance of this model is based on the fact that it provides a nice example of what is called {\em pseudo-hermitian quantum mechanics} (PHQM) in the sense discussed in \cite{mosta,mosta2,bender} and references therein. This is an interesting approach in which the role of self-adjoint operators is replaced by operators satisfying certain rules with respect to the parity and the time reversal operators and, as a consequences, possess eigenvalues which are real or which come in conjugate pairs. However, \cite{tri} neglects many mathematical details of the model, making most of its results  purely formal. Here we discuss a similar model, with the same starting point, but we focus the attention on all those results which can be rigorously proven, and to the assumptions which are needed to prove these results. The bugs in \cite{tri} will be mentioned, and our {\em solutions} will be sketched. In particular, this detailed analysis produces a somehow unexpected result, showing that Riesz bases, \cite{you,chri}, play a crucial role in our context.

 The paper is organized as follows: in the next section we introduce the model and we discuss the Fock states arising from the commutation rules considered. We will show that Riesz bases appear naturally in this context.  In Section III we show how  {\em standard} coherent states (CS), as well as modified CS a la Trifonov, can be introduced. In Section IV we go back to the role of Riesz bases and we discuss our final comments and future projects.

\section{The commutation rules}

Let $\Hil$ be a given Hilbert space with scalar product $\left<.,.\right>$ and related norm $\|.\|$. In \cite{tri} two operators $a$ and $b$ acting on $\Hil$ and satisfying the following commutation rule
\be
[a,b]=\1
\label{21}
\en
were introduced. Of course, this collapses into the CCR if $b=a^\dagger$. It is well known that, exactly because of this rule, $a$ and $b$ cannot both be bounded operators. This simple consideration was just missing in \cite{tri}. Hence we should be careful in dealing with $a$ and $b$ because they cannot be defined in all of $\Hil$. For this reason  we consider the following

\vspace{2mm}

{\bf Assumption 1.--} there exists a non-zero $\varphi_0\in\Hil$ such that $a\varphi_0=0$ and $\varphi_0\in D^\infty(b):=\cap_{k\geq0}D(b^k)$.

\vspace{2mm}

In other words, $\varphi_0$ is annihilated by $a$ and belongs to the domain of all the powers of $b$. Examples of such a vector will be given below. Under this assumption we can introduce the following vectors
\be
\varphi_n=\frac{1}{\sqrt{n!}}\,b^n\,\varphi_0, \quad n\geq 0, \quad \mbox{or}\quad \varphi_n=\frac{1}{\sqrt{n}}\,b\,\varphi_{n-1}, \quad n\geq 1,
\label{22}\en
which clearly belong to $\Hil$ for all $n\geq 0$. Let us now define the  unbounded operator $N:=ba$. Notice that $N\neq N^\dagger$. It is possible to check that $\varphi_n$ belongs to the domain of $N$, $D(N)$, for all $n\geq0$, and that
\be
N\varphi_n=n\varphi_n, \quad n\geq0.
\label{23}\en
Let us now take $\N:=N^\dagger=a^\dagger b^\dagger$. The rule in (\ref{21}) also implies that $[N,b]=b$, $[N,a]=-a$, $[\N,a^\dagger]=a^\dagger$, $[\N,b^\dagger]=-b^\dagger$, and moreover, that
\be
[b^\dagger,a^\dagger]=\1,
\label{24}\en
which again coincides with the CCR if $b^\dagger=a$. However, if this is not the case, this and (\ref{21}) are really different from the CCR. It is clear that all the commutators should be considered in the sense of the unbounded operators. To go on we need another assumption which is analogous to the previous one:

\vspace{2mm}

{\bf Assumption 2.--} there exists a non-zero $\Psi_0\in\Hil$ such that $b^\dagger\Psi_0=0$ and $\Psi_0\in D^\infty(a^\dagger):=\cap_{k\geq0}D((a^\dagger)^k)$.

\vspace{2mm}

Under this assumption we can define the following vectors
\be
\Psi_n=\frac{1}{\sqrt{n!}}(a^\dagger)^n\Psi_0, \quad n\geq 0, \quad \mbox{or}\quad \Psi_n=\frac{1}{\sqrt{n}}(a^\dagger)\Psi_{n-1}, \quad n\geq 1,
\label{25}\en
which clearly belong to $\Hil$ for all $n\geq 0$, and check that they also belong to the domain of $\N$ and that
\be
\N\Psi_n=n\Psi_n, \quad n\geq0.
\label{26}
\en
Incidentally we notice that equation (\ref{26}) implies that $\N$ is unbounded, as well as $N$.
\vspace{2mm}

{\bf Example 1:} this first example shows that the above assumptions need not to be satisfied for generic operators $a$ and $b$. Let $\Hil=\Lc^2(\Bbb{R},d\nu(x))$, where $d\nu(x)=\frac{dx}{1+x^2}$, and let $a=ip$ and $b=x$, where $x$ and $p$ are the quantum position and momentum operators. Then $a\varphi_0(x)=0$ implies that $\varphi_0(x)$ is constant. Of course $\varphi_0(x)\in\Hil$ but $b\varphi_0(x)=x\varphi_0(x)\notin\Hil$. Hence $\varphi_0(x)$ does not belong to $D^\infty(b)$ and Assumption 1 is violated.

\vspace{2mm}

{\bf Example 2:} the second example is that of the harmonic oscillator. In this case $\Hil=\Lc^2(\Bbb{R},dx)$, and taking $a=c:=\frac{1}{\sqrt{2}}\left(\frac{d}{dx}+x\right)$ and $b=c^\dagger=\frac{1}{\sqrt{2}}\left(-\frac{d}{dx}+x\right)$, $[a,b]=[c,c^\dagger]=\1$, we find that $\varphi_0(x)=\Psi_0(x)=\frac{1}{\pi^{1/4}}e^{-x^2/2}$, which satisfies both Assumptions 1 and 2.

\vspace{2mm}

{\bf Example 3:} in this third example, \cite{tri}, we put $\Hil=\Lc^2(\Bbb{R},dx)$ $a_s=c+sc^\dagger$ and $b_s=sc+(1+s^2)c^\dagger$. Hence $[a_s,b_s]=\1$ for all real $s$. Then $a_s\varphi_0(x)=0$ implies that $\varphi_0(x)=N_s\exp\left\{-\frac{1}{2}\,\frac{1+s}{1-s}\,x^2\right\}$, while $b_s^\dagger\Psi_0(x)=0$ is solved by $\Psi_0(x)=N'_s\exp\left\{-\frac{1}{2}\,\frac{1+s+s^2}{1-s+s^2}\,x^2\right\}$.
Here $N_s$ and $N'_s$ are $s-$depending normalization constants. Of course, in order for both these functions to be square integrable we should require that both $\frac{1+s}{1-s}$ and $\frac{1+s+s^2}{1-s+s^2}$ are positive, which is true if $-1<s<1$. This same condition ensures also that $\varphi_0(x)\in D^\infty(b_s)$ and that $\Psi_0(x)\in D^\infty(a_s^\dagger)$: any polynomial multiplied for a gaussian function belongs to $\Lc^2(\Bbb{R},dx)$.

A minor modification of this example is also discussed in \cite{tri}: again we have $\Hil=\Lc^2(\Bbb{R},dx)$ and $a_s=c+sc^\dagger$, but we choose $b_s=-sc+(1-s^2)c^\dagger$. Hence $[a_s,b_s]=\1$ for all real $s$, $\varphi_0(x)$ is the same as above while $\Psi_0(x)=N''_s\exp\left\{-\frac{1}{2}\,\frac{1-s-s^2}{1+s-s^2}\,x^2\right\}$. The main difference with respect to the previous case is in the range of $s$ in which Assumptions 1 and 2 are satisfied: we need now to restrict $s$ in the interval $\left(\frac{1}{2}(1-\sqrt{5}),\frac{1}{2}(-1+\sqrt{5})\right)$.

\vspace{2mm}

{\bf Example 4:} in the previous example $a$ and $b$ are defined by introducing a one-dimensional {\em deformation parameter} $s$ starting from the bosonic operators $c$ and $c^\dagger$. Now we generalize this procedure, showing that also two-dimensional deformations are allowed. Let $a_{\alpha,\mu}:=\alpha c+\frac{\alpha}{\mu} c^\dagger$, $b_{\alpha,\mu}:=\mu\frac{\alpha^2-1}{\alpha} c+\alpha c^\dagger$, where $\alpha$ and $\mu$ are real constants such that $\alpha,\mu\neq0$ and $\alpha^2\neq\mu^2(\alpha^2-1)$. Hence $a_{\alpha,\mu}^\dagger\neq b_{\alpha,\mu}$ (which would trivialize the situation), and $[a_{\alpha,\mu},b_{\alpha,\mu}]=\1$. The solutions of $a_{\alpha,\mu}\varphi_0(x)=0$ and $b_{\alpha,\mu}^\dagger\Psi_0(x)=0$ are respectively $\varphi_0(x)=N_{\alpha,\mu}\exp\left\{-\frac{1}{2}\,\frac{\mu+1}{\mu-1}\,x^2\right\}$, and $\Psi_0(x)=N'_{\alpha,\mu}\exp\left\{-\frac{1}{2}\,\frac{\alpha^2+\mu(\alpha^2-1)}{\alpha^2-\mu(\alpha^2-1)}\,x^2\right\}$.
Again, $N_{\alpha,\mu}$ and $N'_{\alpha,\mu}$ are normalization constants. For these functions to satisfy  Assumptions 1 and 2 it is enough to have  $\frac{\mu+1}{\mu-1}>0$ and $\frac{\alpha^2+\mu(\alpha^2-1)}{\alpha^2-\mu(\alpha^2-1)}>0$, which are both verified if we take $\alpha>1$ and $1<\mu<1+\frac{1}{\alpha^2-1}$.

\vspace{2mm}

In the above assumptions it is now easy to check that $\left<\Psi_n,\varphi_m\right>=\delta_{n,m}\left<\Psi_0,\varphi_0\right>$ for all $n, m\geq0$, which, if we take $\Psi_0$ and $\varphi_0$ such that $\left<\Psi_0,\varphi_0\right>=1$, becomes
\be
\left<\Psi_n,\varphi_m\right>=\delta_{n,m}, \quad \forall n,m\geq0
\label{27}\en
This means that the $\Psi_n$'s and the $\varphi_n$'s are biorthogonal. It is also possible to prove the following Lemma, which will be useful in the rest of the paper
\begin{lemma}
For all $n\geq0$ we have  $\varphi_n\in D(a)$ and $\Psi_n\in D(b^\dagger)$. Moreover
\be
a\varphi_n=\left\{
    \begin{array}{ll}
        0,\hspace{3.9cm}\mbox{ if } n=0,  \\
        \sqrt{n}\,\varphi_{n-1}, \hspace{2.6cm} \mbox{ if } n>0,\\
       \end{array}
        \right.
\label{28}\en
and
\be
b^\dagger\Psi_n=\left\{
    \begin{array}{ll}
        0,\hspace{3.9cm}\mbox{ if } n=0,  \\
        \sqrt{n}\,\Psi_{n-1}, \hspace{2.6cm} \mbox{ if } n>0.\\
       \end{array}
        \right.
\label{29}\en
\end{lemma}

Once again, the proof is simple and will not be given here. It is maybe more relevant to remark that, since the vectors in $\F_\varphi:=\{\varphi_n,\,n\geq0\}$ and in $\F_\Psi:=\{\Psi_n,\,n\geq0\}$ are not orthogonal ($\left<\varphi_n,\varphi_k\right>\neq \delta_{n,k}$ and $\left<\Psi_n,\Psi_k\right>\neq \delta_{n,k}$ in general), hence equation (\ref{28})  does not automatically imply that $a^\dagger\varphi_n=\sqrt{n+1}\,\varphi_{n+1}$, as it would do if $\F_\varphi$ were an orthonormal (o.n.) basis of $\Hil$. For the same reason $b\,\Psi_n\neq\sqrt{n+1}\,\Psi_{n+1}$, in general. However, the sets $\F_\varphi$ and $\F_\Psi$ are biorthogonal and, because of this, the vectors of each set are linearly independent. If we now call $\D_\varphi$ and $\D_\Psi$ respectively the linear span of  $\F_\varphi$ and $\F_\Psi$, and $\Hil_\varphi$ and $\Hil_\Psi$ their closures, by construction $\F_\varphi$ is complete in $\Hil_\varphi$ and $\F_\Psi$ is complete in $\Hil_\Psi$. More than this, we can also prove that
\be
f=\sum_{n=0}^\infty \left<\Psi_n,f\right>\,\varphi_n, \quad \forall f\in\Hil_\varphi,\qquad  h=\sum_{n=0}^\infty \left<\varphi_n,f\right>\,\Psi_n, \quad \forall h\in\Hil_\Psi
\label{210}\en
What is not in general ensured is that the Hilbert spaces introduced so far all coincide, i.e. that $\Hil_\varphi=\Hil_\Psi=\Hil$. With our assumptions we can only state that $\Hil_\varphi\subseteq\Hil$ and $\Hil_\Psi\subseteq\Hil$. However, in all the examples considered previously, these three Hilbert spaces really coincide and for this reason we also consider the following

\vspace{2mm}

{\bf Assumption 3.--} The above Hilbert spaces all coincide: $\Hil_\varphi=\Hil_\Psi=\Hil$.

\vspace{2mm}

We would like to mention that this problem was not considered in \cite{tri}, where it was taken for granted. From (\ref{210}) we deduce, first of all, that both $\F_\varphi$ and $\F_\Psi$ are bases in $\Hil$. The resolution of the identity looks now
\be
\sum_{n=0}^\infty
|\varphi_n><\Psi_n|=\sum_{n=0}^\infty
|\Psi_n><\varphi_n|=\1,
\label{211}\en
where $\1$ is the common identity of all the Hilbert spaces and where the useful Dirac bra-ket notation has been adopted. At this stage it is natural to introduce two operators which we now write formally using again the bra-ket notation as
\be
\eta_\varphi=\sum_{n=0}^\infty
|\varphi_n><\varphi_n|,\qquad \eta_\Psi=\sum_{n=0}^\infty
|\Psi_n><\Psi_n|.
\label{212}\en
Of course, these operators need not to be well defined: for instance the series could  be not convergent, or even if they do, they could converge to some unbounded operator, so we have to be careful about domains. Again, this aspect was not considered in \cite{tri}.

Let us then introduce, more rigorously, an operator $\eta_\varphi$ acting on a vector $f$ in its domain $D(\eta_\varphi)$ as $\eta_\varphi f=\sum_{n=0}^\infty\left<\varphi_n,f\right>\varphi_n$. We also introduce a second operator, $\eta_\Psi$, acting on a vector $h$ in its domain $D(\eta_\Psi)$ as $\eta_\Psi h=\sum_{n=0}^\infty\left<\Psi_n,h\right>\Psi_n$. Under Assumption 3, both these operators are densely defined in $\Hil$ since $\D_\varphi\subseteq D(\eta_\varphi)$ and $\D_\Psi\subseteq D(\eta_\Psi)$. In particular, we find that
\be
\eta_\varphi\Psi_n=\varphi_n,\qquad
\eta_\Psi\varphi_n=\Psi_n,
\label{213}\en
for all $n\geq0$, which also implies that $\Psi_n=(\eta_\Psi\eta_\varphi)\Psi_n$ and $\varphi_n=(\eta_\varphi\eta_\Psi)\varphi_n$, for all $n\geq0$. Hence
\be
\eta_\Psi\eta_\varphi=\eta_\varphi\eta_\Psi=\1 \quad \Rightarrow \quad \eta_\Psi=\eta_\varphi^{-1}.
\label{214}\en
In other words, both $\eta_\Psi$ and $\eta_\varphi$ are invertible and one is the inverse of the other. Furthermore, we can also check that they are both positive defined and symmetric. One may wonder whether they are automatically bounded, then. This is not so, in general. Indeed it is not hard to construct examples of unbounded positive and symmetric operators mapping a basis of $\Hil$ in a biorthogonal basis. It is enough to consider a number operator $\hat N$ defined on an o.n. basis of $\Hil$, $\{e_n,\, n\geq 1\}$, as $\hat N\,e_n=ne_n$, $n\geq 1$. Hence, calling $\varphi_n=\frac{1}{\sqrt{n}}\,e_n$ and $\Psi_n=\sqrt{n}\,e_n$, $\F_\varphi$ and $\F_\Psi$ are biorthogonal  bases of $\Hil$. Moreover $\hat N\varphi_n=\Psi_n$, $\hat N^{-1}\Psi_n=\varphi_n$, $\hat N>0$, but $\hat N$ is unbounded. A simple modification of example also provides an example in which the positive operator $M_{1\rightarrow2}$ mapping a basis $\G_1$ in its biorthogonal basis $\G_2$ and its inverse $M_{2\rightarrow1}=M_{1\rightarrow2}^{-1}$ mapping $\G_2$ into $\G_1$, are both unbounded. For that it is enough to define
$$
\varphi_n=\left\{
    \begin{array}{ll}
        \frac{1}{n}\,e_n,\hspace{.1cm}\mbox{ if $n$ is even},  \\
        n\,e_n,\hspace{.1cm}\mbox{ if $n$ is odd},\\
       \end{array}
        \right.
        \Psi_n=\left\{
    \begin{array}{ll}
        n\,e_n,\hspace{.1cm}\mbox{ if $n$ is even},  \\
        \frac{1}{n}\,e_n,\hspace{.1cm}\mbox{ if $n$ is odd},\\
       \end{array}
        \right. \mbox{ and }
     M_{1\rightarrow2}\,e_n=\left\{
    \begin{array}{ll}
        n^2\,e_n,\hspace{.1cm}\mbox{ if $n$ is even},  \\
        \frac{1}{n^2}\,e_n,\hspace{.1cm}\mbox{ if $n$ is odd}.\\
       \end{array}
        \right.
$$

This is not a big surprise because, as discussed in \cite{you}, two biorthogonal bases are related by a bounded operator, with bounded inverse, if and only if they are Riesz basis. This suggests we need some more assumption to go further. In order to keep the roles of $\F_\varphi$ and $\F_\Psi$ symmetric we require now  the following

\vspace{2mm}

{\bf Assumption 4.--} $\F_\varphi$ and $\F_\Psi$ are Bessel sequences. In other words, there exist two positive constants $A_\varphi,A_\Psi>0$ such that, for all $f\in\Hil$,
\be
\sum_{n=0}^\infty\,|\left<\varphi_n,f\right>|^2\leq A_\varphi\,\|f\|^2,\qquad \sum_{n=0}^\infty\,|\left<\Psi_n,f\right>|^2\leq A_\Psi\,\|f\|^2.
\label{215}\en

\vspace{2mm}

As a consequence, see \cite{chri}, $\F_\varphi$ and $\F_\Psi$ are both Riesz bases with bounds $\left(\frac{1}{A_\Psi},A_\varphi\right)$ and $\left(\frac{1}{A_\varphi},A_\Psi\right)$. This is due to the fact that they are biorthogonal sets. In particular then, $\F_\varphi$ and $\F_\Psi$ are frames.

\begin{lemma}
Under Assumption 4 both $\eta_\varphi$ and $\eta_\Psi$ are bounded operators. In particular we have
\be
\|\eta_\varphi\|\leq A_\varphi, \qquad \|\eta_\Psi\|\leq A_\Psi.
\label{216}\en Moreover
\be
\frac{1}{A_\Psi}\,\1\leq \eta_\varphi \leq A_\varphi\,\1,\qquad \frac{1}{A_\varphi}\,\1\leq \eta_\Psi \leq A_\Psi\,\1.
\label{217}\en
\end{lemma}
\begin{proof}
We just prove that $\|\eta_\varphi\|\leq A_\varphi$. Using (\ref{215}), (\ref{212}) and the Schwarz inequality we have
$$
\|\eta_\varphi\|=\sup_{\|f\|=\|g\|=1}\left|\left<f,\eta_\varphi g\right>\right|=
\sup_{\|f\|=\|g\|=1}\left|\sum_{k=0}^\infty\left<f,\varphi_k\right>\left< \varphi_k,g\right>\right|\leq $$ $$\leq\sup_{\|f\|=\|g\|=1} \sqrt{\sum_{k=0}^\infty\left|\left<f,\varphi_k\right>\right|^2}
\sqrt{\sum_{k=0}^\infty\left|\left<\varphi_k,g\right>\right|^2}\leq A_\varphi\,\sup_{\|f\|=\|g\|=1}\|f\|\,\|g\|        \leq A_\varphi
$$

\end{proof}

This Lemma implies that the domains of $\eta_\varphi$ and $\eta_\Psi$ can be taken to be all of $\Hil$. An interesting consequence of our construction is the following

\begin{cor}
The set $\F_\Psi$ coincides with the dual frame $\tilde\F_\varphi$ of $\F_\varphi$. Also, $\F_\varphi$ coincides with the dual frame $\tilde\F_\Psi$ of $\F_\Psi$.
\end{cor}
\begin{proof}
Since $\F_\varphi$ is a frame, its frame operator  $S_\varphi$, defined as $S_\varphi f=\sum_{n=0}^\infty \left<\varphi_n,f\right>\varphi_n$, $f\in\Hil$, is well defined, bounded and invertible. It is also clear that it coincides with $\eta_\varphi$. Hence, recalling that the vectors $\tilde\varphi_n$ of the  set $\tilde\F_\varphi$ are defined as  $\tilde\varphi_n=S_\varphi^{-1}\varphi_n$, we have
$$
\tilde\varphi_n=S_\varphi^{-1}\varphi_n=\eta_\varphi^{-1}\varphi_n=\eta_\Psi\varphi_n=\Psi_n,
$$
where we have used (\ref{213}) and (\ref{214}). Our second assertion can be proved similarly.
\end{proof}

\subsection{Connections with intertwining operators}

Under the assumptions we have considered so far a natural structure appears in which $\eta_\Psi$ and $\eta_\varphi$ play the role of intertwining operators between non self-adjoint operators. This looks interesting since in the literature on the subject, \cite{intop,bag1,bag2,bag3} and references therein, intertwining operators usually act between self-adjoint operators {\em preserving} the spectra and modifying the eigenvectors quite easily. More explicitly, suppose that $h_1$ and $h_2$ are two self-adjoint operators and that a third operator $x$ satisfies the  intertwining equation $xh_1=h_2x$. Let $e_n^{(1)}$ be an eigenstate of $h_1$ with eigenvalue $\epsilon_n$, $h_1\,e_n^{(1)}=\epsilon_n\,e_n^{(1)}$. Now, if $e_n^{(2)}:=xe_n^{(1)}\neq 0$, then we can check with a straight computation that $h_2e_n^{(2)}=\epsilon_n\,e_n^{(2)}$.

We begin our analysis remarking that a simple use of induction on $n$ proves that for all $n\geq 0$ the vector $\varphi_n$ belongs to the domain of the operator $\eta_\varphi\,a^\dagger\,\eta_\Psi$ and that
\be
b\varphi_n=\eta_\varphi\,a^\dagger\,\eta_\Psi\varphi_n,
\label{218}\en
which can also be written, recalling that $\varphi_n=\eta_\varphi\Psi_n$ and that $\eta_\Psi\varphi_n=\Psi_n$, as $b\,\eta_\varphi\Psi_n=\eta_\varphi\,a^\dagger\Psi_n$ or yet, as $b\,\eta_\varphi=\eta_\varphi\,a^\dagger$. Hence $\eta_\varphi$ intertwines between $b$ and $a^\dagger$. Analogously we can prove that $a^\dagger \eta_\Psi=\eta_\Psi\,b$, while other intertwining relations can be found just taking the adjoint of these equations. Two interesting consequences are now the following equalities
\be
\eta_\Psi\,N=\N\eta_\Psi \quad \mbox{ and }\quad N\,\eta_\varphi=\eta_\varphi\,\N,
\label{219}\en
whose proof is straightforward. It is now trivial to check that our results are coherent with the standard technique of intertwining operators, but for the lack of self-adjointness of the operators $N$ and $\N$. This is suggested by the fact that $N$ and $\N$ have the same eigenvalues. More in details, using for instance the intertwining relation $N\,\eta_\varphi=\eta_\varphi\,\N$, together with the equations $\varphi_n=\eta_\varphi\Psi_n$ and $\N\Psi_n=n\Psi_n$, we can easily check that $\varphi_n$ are eigenstates of $N$ with eigenvalue $n$. The computation goes as follows:
$$
N\varphi_n=N\,\eta_\varphi\Psi_n=\eta_\varphi\,\N\,\Psi_n=\eta_\varphi\,(n\,\Psi_n)=n\,\varphi_n.
$$
It is also worth noticing that condition (\ref{219}) is a pseudo-hermiticity condition for the operators $N$ and $\N$. Indeed we have $\eta_\Psi\,N\,\eta_\Psi^{-1}=N^\dagger$ and $\eta_\varphi\,N^\dagger\,\eta_\varphi^{-1}=N$, which because of the properties of $\eta_\Psi$ and $\eta_\varphi$, are exactly the conditions which state that $N$ and $\N$ are pseudo-hermitian conjugate, \cite{mosta}. We recall that this was just the main motivation in \cite{tri} for considering the  commutation rules in (\ref{21}).

\subsection{Inverting the construction}

What we have discussed so far shows in particular that, under Assumptions 1-4, $b$ and $a^\dagger$ are necessarily related by $\eta_\varphi$ as in (\ref{218}) or, equivalently, as in $b=\eta_\varphi\,a^\dagger\,\eta_\varphi^{-1}$. We are now interested in considering the {\em inverse construction}, i.e. in considering as our starting point again two operators $a$ and $b$ satisfying $[a,b]=\1$ under the assumption that $b$ is related to $a^\dagger$ in the way shown above, and check what happens.

Let therefore $a$ be a given operator defined on a dense domain of a given Hilbert space, $D(a)\subseteq\Hil$, with adjoint $a^\dagger$  densely defined. We now consider a bounded, positive, operator $T$, with bounded inverse such that a dense subset of $\Hil$ exists, $\E$, with $T^{-1}:\E\rightarrow D(a^\dagger)$. Hence the operator $b_T:=T\,a^\dagger\,T^{-1}$ is densely defined since $\E\subseteq D(b_T)$. We assume that
\be
[a,b_T]=\1
\label{220}\en
\vspace{2mm}

{\bf Remark:--} If we work in the Assumptions 1-4 above, taking $T$ as the frame operator of $\F_\varphi$ we get an example of this settings.

\vspace{2mm}

As before, we need to extract a certain set of conditions if we want to deduce some interesting results. The first assumption is exactly Assumption 1 above, which we now write as

\vspace{2mm}

{\bf Assumption I.--} there exists a non-zero $\varphi_0\in\Hil$ such that $a\varphi_0=0$ and $T^{-1}\varphi_0\in D^\infty(a^\dagger)$.

\vspace{2mm}

Hence (\ref{22}) can be extended also to this settings and we have
\be
\varphi_n=\frac{1}{\sqrt{n!}}\,b_T^n\,\varphi_0, \quad n\geq 0, \quad \mbox{or}\quad \varphi_n=\frac{1}{\sqrt{n}}\,b_T\,\varphi_{n-1}, \quad n\geq 1,
\label{221}\en
which are in the domain  of the operator $N_T:=b_T\,a$ and satisfy the eigenvalue equation  $N_T\varphi_n=n\,\varphi_n$, for all $n\geq 0$. As before we define $\N_T:=N_T^\dagger=a^\dagger\,b_T^\dagger$. There is no need of require here the analogous of Assumption 2. Indeed, if we define $\Psi_0:=T^{-1}\varphi_0$, it is first clear that $b_T^\dagger\Psi_0=0$. Moreover, due to Assumption I above, $\Psi_0\in D^\infty(a^\dagger)$. Hence the vectors
$$
\Psi_n=\frac{1}{\sqrt{n!}}(a^\dagger)^n\Psi_0, \quad n\geq 0, \quad \mbox{or}\quad \Psi_n=\frac{1}{\sqrt{n}}(a^\dagger)\Psi_{n-1}, \quad n\geq 1
$$
are well defined, belong to the domain of $\N_T$, satisfy the eigenvalue equation $\N_T\Psi_n=n\Psi_n$, $n\geq0$, and the following relation holds
\be
\varphi_n=T\Psi_n,\qquad \Psi_n=T^{-1}\varphi_n,
\label{222}\en
for all $n\geq0$. Notice that these look exactly like the equations in (\ref{213}). Moreover, if $\left<\Psi_0,\varphi_0\right>=1$, then $\left<\Psi_n,\varphi_m\right>=\delta_{n,m}$. From this biorthogonality condition  two important estimates on $\|\varphi_n\|$ and $\|\Psi_n\|$ can be deduced. For instance, since for all $n\geq0$
$$
1=\left<\Psi_n,\varphi_n\right>=\left<T^{-1}\varphi_n,\varphi_n\right>=\|T^{-1/2}\varphi_n\|^2,$$
we deduce that
\be
\|\varphi_n\|=\|T^{1/2}\,T^{-1/2}\varphi_n\|\leq \|T^{1/2}\|.
\label{223}\en
Analogously we can prove that $\|\Psi_n\|\leq \|T^{-1/2}\|$, $\forall n\geq0$. Defining $\F_\varphi$, $\F_\Psi$, $\Hil_\varphi$ and $\Hil_\Psi$ as before, we consider now the following:

\vspace{2mm}

{\bf Assumption II.--} The above Hilbert spaces all coincide: $\Hil_\varphi=\Hil_\Psi=\Hil$.

\vspace{2mm}
This is exactly our previous Assumption 3. Hence, \cite{you}, since $\F_\varphi$ and $\F_\Psi$ are two (biorthogonal) bases of $\Hil$ related by a bounded operator $T$ with bounded inverse, $\F_\varphi$ are $\F_\Psi$ are necessarily Riesz bases. Moreover, defining $\eta_\varphi$ and $\eta_\Psi$ as in (\ref{212}), it is easy to check that they coincide with $T$ and $T^{-1}$, so that they are bounded operators with bounded inverse, mapping $\F_\Psi$ into $\F_\varphi$ and vice-versa. Moreover, $\F_\varphi$ and $\F_\Psi$ are dual frames of each other. Finally, $\eta_\varphi$ and $\eta_\Psi$ are the frame operators respectively of $\F_\varphi$ and $\F_\Psi$. Hence essentially the same general structure discussed previously is recovered.

\vspace{2mm}

We will continue with this analysis in the last section, where other aspects and applications of Riesz bases in this context will be considered.

\section{Coherent states}

In \cite{tri} a family of CS for the model has been introduced. Again, in our opinion some more mathematical care is required. For this reason we carry on our own analysis, focusing the attention on those points which may create some problems. Our analysis is also motivated by the work in \cite{ali}, where the role of non-orthogonal bases in the description of coherent states is discussed.

We work here under Assumptions 1-4 of the previous section. Hence there exist $\varphi_0$ and $\Psi_0$ in $\Hil$ such that $a\varphi_0=b^\dagger\Psi_0=0$. Also, $\varphi_0\in D^\infty(b)$ and $\Psi_0\in D^\infty(a^\dagger)$. Let us introduce the $z$-dependent operators
\be
U(z)=\exp\{z\,b-\overline{z}\,a\}, \qquad V(z)=\exp\{z\,a^\dagger-\overline{z}\,b^\dagger\},
\label{31}\en
$z\in\E\subseteq\Bbb{C}$ to be identified, and the following vectors:
\be
\varphi(z)=U(z)\varphi_0,\qquad \Psi(z)=V(z)\,\Psi_0.
\label{32}\en
It is possible to check that these vectors are well defined in $\Hil$ for all $z\in\Bbb{C}$. This can be checked using the Baker-Campbell-Hausdorff formula which produces here the identities
$$
U(z)=e^{-|z|^2/2}\,e^{z\,b}\,e^{-\overline{z}\,a},\qquad V(z)=e^{-|z|^2/2}\,e^{z\,a^\dagger}\,e^{-\overline{z}\,b^\dagger},
$$
together with the properties of $\varphi_0$ and $\Psi_0$. We get
\be
\varphi(z)=e^{-|z|^2/2}\,\sum_{n=0}^\infty\,\frac{z^n}{\sqrt{n!}}\,\varphi_n, \qquad
\Psi(z)=e^{-|z|^2/2}\,\sum_{n=0}^\infty\,\frac{z^n}{\sqrt{n!}}\,\Psi_n.
\label{33}\en
However, since $U(z)$ and $V(z)$ are not unitary operators, or alternatively since $\varphi_n$ and $\Psi_n$ are not normalized in general, we should check that these series both converge. For that it is convenient to repeat the same steps which have produced under different assumptions the estimate in (\ref{223}). We get easily $\|\varphi_n\|\leq \|\eta_\Psi^{-1/2}\|$ and $\|\Psi_n\|\leq \|\eta_\varphi^{-1/2}\|$, for all $n\geq0$. Hence the series in (\ref{33}) are both norm convergent for all possible $z\in\Bbb{C}$. These vectors are called {\em coherent} since they are eigenstates of some lowering operators. Indeed we can check that
\be
a\varphi(z)=z\varphi(z), \qquad b^\dagger\Psi(z)=z\Psi(z),
\label{34}\en
for all $z\in\Bbb{C}$. It is also a standard exercise, putting $z=r\,e^{i\theta}$, to check the following operator equalities:
\be
\frac{1}{\pi}\int_{\Bbb{C}}\,dz |\varphi(z)><\varphi(z)|=\eta_\varphi, \qquad
\frac{1}{\pi}\int_{\Bbb{C}}\,dz |\Psi(z)><\Psi(z)|=\eta_\Psi,
\label{35}\en
as well as
\be
\frac{1}{\pi}\int_{\Bbb{C}}\,dz |\varphi(z)><\Psi(z)|=
\frac{1}{\pi}\int_{\Bbb{C}}\,dz |\Psi(z)><\varphi(z)|=\1,
\label{36}\en
which are written in convenient bra-ket notation.
This last equality was formally derived in \cite{tri}, where no analysis on the convergence of the series defining the  CS was considered.

A natural question to ask when dealing with CS is whether some kind of Heisenberg uncertainty relation is saturated. But the natural operator replacing the position operator $\frac{1}{\sqrt{2}}(a+a^\dagger)$ in our context should be $\frac{1}{\sqrt{2}}(a+b)$, which is no longer self-adjoint. Hence the problem is not necessarily well defined, on these states and with these operators. We will show how this problem can be reconsidered below.

\vspace{3mm}

An interesting feature of our system is the following: under the Assumptions 1-4 it is clear that the set $\F_{\hat\varphi}:=\{\hat\varphi_n=S_\varphi^{-1/2}\varphi_n\}$ is an o.n. basis of $\Hil$. This means that, defining
\be
\hat\varphi(z)=e^{-|z|^2/2}\,\sum_{n=0}^\infty\,\frac{z^n}{\sqrt{n!}}\,\hat\varphi_n,
\label{37}\en
these are  {\em standard} CS, \cite{kla}. In particular they are normalized, $\left<\hat\varphi(z),\hat\varphi(z)\right>=1$ for all $z\in\Bbb{C}$,  and satisfy the following resolution of the identity: $\frac{1}{\pi}\int_{\Bbb{C}}\,dz |\hat\varphi(z)><\hat\varphi(z)|=\1$. Also,  they are related to the states in (\ref{32}) as follows: $\varphi(z)=S_\varphi^{1/2}\hat\varphi(z)$, for all $z\in\Bbb{C}$.

Moreover, if we define the new operator $a_\varphi:=S_\varphi^{-1/2}\,a\,S_\varphi^{1/2}$, we also deduce that
\be
a_\varphi\hat\varphi(z)=z\hat\varphi(z),
\label{38}\en
which shows that $\hat\varphi(z)$ are eigestates of a certain operator $a_\varphi$ (related to $a$ and to the structure of the system) with eigenvalue $z$. The action of the operator $a_\varphi$ on the o.n. basis $\F_{\hat\varphi}$ is given by
\be
a_\varphi\hat\varphi_n=\left\{
    \begin{array}{ll}
        0,\hspace{3.9cm}\mbox{ if } n=0,  \\
        \sqrt{n}\,\hat\varphi_{n-1}, \hspace{2.6cm} \mbox{ if } n>0,\\
       \end{array}
        \right.
\label{39}\en
which implies that $a_\varphi^\dagger\hat\varphi_n=\sqrt{n+1}\,\hat\varphi_{n+1}$, $n\geq0$, and, therefore, that $[a_\varphi,a_\varphi^\dagger]=\1$. We stress that this does not means that $[a,a^\dagger]=\1$ as well, since $S_\varphi$ is not unitary.

Therefore, if we now define the operators $x_\varphi:=\frac{1}{\sqrt{2}}(a_\varphi+a_\varphi^\dagger)$ and $p_\varphi:=\frac{1}{i\,\sqrt{2}}(a_\varphi-a_\varphi^\dagger)$ they are self-adjoint and, with standard notation, we get $\Delta x_\varphi\,\Delta p_\varphi=\frac{1}{2}$. So the $\hat\varphi(z)$'s saturate the Heisenberg relation, even if for the {\em original} $\varphi(z)$'s  it was not even clear which operators we had to consider.

\vspace{3mm}

Of course, what we have done above starting from $\F_\varphi$, could be repeated starting from $\F_\Psi$. So we use its frame operator $S_\Psi$, which coincides with $\eta_\Psi$ in our hypotheses, to define the following o.n. basis of $\Hil$: $\F_{\hat\Psi}:=\{\hat\Psi_n=S_\varphi^{1/2}\Psi_n\}$. However, this does not produce new results. Indeed we find that $\F_{\hat\varphi}=\F_{\hat\Psi}$ since
$$
\hat\Psi_n=S_\varphi^{1/2}\Psi_n=S_\varphi^{1/2}\eta_\Psi\varphi_n=S_\Psi^{1/2}\varphi_n=\hat\varphi_n,
$$
where we have used, in particular, the relations between the frame operators and the operator $\eta_\Psi$.

It is finally possible to find a self-adjoint operator $N_\varphi$ whose eigenstates are exactly the vectors in $\F_{\hat\varphi}$. Again, the strategy is to use  the frame operators $S_\varphi$ and $S_\Psi$: if we put $N_\varphi:=S_\varphi^{-1/2}\,N\,S_\varphi^{1/2}$ and $\N_\Psi:=S_\Psi^{-1/2}\,\N\,S_\Psi^{1/2}$, it is possible to check that $N_\varphi=\N_\Psi=a_\varphi^\dagger\,a_\varphi$ and that $N_\varphi\hat\varphi_n=n\hat\varphi_n$ for all $n\geq0$.

\vspace{2mm}

{\bf Remark:--} Of course, having the o.n. basis $\F_{\hat\varphi}$, we could use them to construct several different kind of CS, like the non-linear ones, \cite{alibag} and references therein, or the Gazeau-Klauder CS, \cite{gk}. The first ones mainly differ from the $\hat\varphi(z)$ we have introduced here since the sequence $\{n\}$ is replaced by a different set, $\{\epsilon_n\}$, of positive numbers with $\epsilon_0=0$ so that they are not defined, in general, in all of $\Bbb{C}$ but only in a certain {\em domain of convergence}. The second ones are built up from a given hamiltonian, which could be the operator $N_\varphi$ above, or some different self-adjoint operator with positive increasing eigenvalues.

\vspace{2mm}

We conclude that, under Assumptions 1-4 above, we can introduce different kind of CS, with similar properties. Once again, our treatment displays the relevance of the Riesz bases in the present context.

\section{The role of Riesz bases and conclusions}

We have constructed and discussed in details a physical example which extends the CCR  illustrating Theorem 6.1.1 of \cite{chri}. We have also discussed that this example is strongly related to the theory of Riesz bases and frames and that it can be used as an example of the theory of intertwining operators where the so called {\em hamiltonians} are not necessarily self-adjoint.

We have also used this model to discuss some properties of CS arising from non o.n. bases.

We end the paper showing that Riesz bases really play a crucial role in our analysis, and in particular that to any Riesz basis we can associate two operators $a$ and $b$ satisfying $[a,b]=\1$ and for which Assumptions 1-4 of Section II are satisfied. Hence each Riesz basis produce a concrete example of our framework.

Let $\F_\varphi:=\{\varphi_n,\,n\geq0\}$ be a Riesz basis of $\Hil$ with bounds $A$ and $B$, $0<A\leq B<\infty$. The associated frame operator $S:=\sum_{n=0}^\infty\,|\varphi_n><\varphi_n|$ is bounded, positive and admits a bounded inverse. The set $\F_{\hat\varphi}:=\{\hat\varphi_n:=S^{-1/2}\varphi_n,\,n\geq0\}$ already introduced in the previous section is an o.n. basis of $\Hil$. Hence we can define a lowering operator $a_\varphi$ on  $\F_{\hat\varphi}$ as in (\ref{39}), and its adjoint, $a_\varphi^\dagger$, as $a_\varphi^\dagger\hat\varphi_n=\sqrt{n+1}\,\hat\varphi_{n+1}$, $n\geq0$. Hence $[a_\varphi,a_\varphi^\dagger]=\1$. If we now define $a:=S^{1/2}\,a_\varphi\,S^{-1/2}$, this acts on $\F_\varphi$ as in (2.8). Hence $a$ is also a lowering operator. However, since $\F_\varphi$ is not an o.n. basis in general, $a^\dagger$ is not a raising operator, contrarily to $a_\varphi^\dagger$. Hence $[a,a^\dagger]\neq\1$. If we now define the operator $b:=S^{1/2}\,a_\varphi^\dagger\,S^{-1/2}$, it is clear that in general  $b\neq a^\dagger$. Moreover, $b$ acts on $\varphi_n$ as a raising operator: $b\,\varphi_n=\sqrt{n+1}\,\varphi_{n+1}$, for all $n\geq0$, and we also have $[a,b]=\1$. So we have constructed two operators satisfying (\ref{21}) and which are not related by a simple conjugation. This is not the end of the story. Indeed:
\begin{enumerate}
\item Assumption 1 is verified since $\varphi_0$ is annihilated by $a$ and belongs to the domain of all the powers of $b$. In particular we find that $b^n\,\varphi_0=\sqrt{n!}\,\varphi_n$, $\forall n\geq 0$.
\item As for Assumption 2, it is enough to define $\Psi_0=S^{-1}\,\varphi_0$. With this definition $b^\dagger\,\Psi_0=0$ and $\Psi_0$ belongs to the domain of all the powers of $a^\dagger$. In particular  $(a^\dagger)^n\,\Psi_0=\sqrt{n!}\,S^{-1/2}\,\hat\varphi_n$, which is a well defined vector in $\Hil$  for all $n\geq 0$.
\item Since $\F_\varphi$ is a Riesz basis of $\Hil$ by assumption, then $\Hil_\varphi=\Hil$. Notice now that the vector $\Psi_n$ in (\ref{25}) can be written as $\Psi_n=S^{-1}\,\varphi_n$, for all $n\geq0$. Hence $\F_\Psi$ is in duality with $\F_\varphi$ and therefore is a Riesz basis of $\Hil$ as well. Hence $\Hil_\Psi=\Hil$. This prove Assumption 3.
\item As for Assumption 4, this is equivalent to the hypothesis originally assumed here, i.e. that $\F_\varphi$ is a Riesz basis.
\end{enumerate}

The results discussed along this paper have an interesting consequence related to PHQM, where the operator $\eta_\varphi$ is called {\em the metric operator} and is used both to define a new scalar product in the Hilbert space of the theory and the so-called pseudo-hermitian conjugate of an operator, \cite{mosta,mosta2}. Indeed, from what we have discussed here, the computation of this metric operator seems to be not very different from the computation of the frame
operator for a certain Riesz basis, for which some perturbative expansions can be found in \cite{dau,bagframes}. We also would like to mention that the role of Riesz basis within PHQM was in part already recognized and discussed in \cite{mosta}.

We could wonder what may change in this construction if we replace a Riesz basis with a frame: the main properties of the frame operator $S$ do not change, in fact: it is still bounded, self-adjoint, and with bounded inverse. However, linear independence of the vectors is lost and the set $\F_{\hat\varphi}$ is no longer a basis. Hence $a_\varphi$ cannot be defined as we have done here. Analogous problems also arise in the construction of CS discussed in the previous section, see also \cite{fmt}. So the conclusion is that for what we had in mind in this paper, Riesz bases are much better than simple frames!

\section*{Acknowledgements}

It is a pleasure to thank Prof. Trapani for many interesting discussions at a preliminary stage of this work.  The author also acknowledges financial support by the Murst, within the  project {\em Problemi
Matematici Non Lineari di Propagazione e Stabilit\`a nei Modelli
del Continuo}, coordinated by Prof. T. Ruggeri.

\end{document}